\documentclass[journal]{IEEEtran}
\IEEEoverridecommandlockouts
\usepackage{cite}
\usepackage{amsmath,amssymb,amsfonts}
\usepackage{graphicx}
\usepackage{textcomp}
\usepackage{xcolor}
\usepackage{multirow}
\usepackage{array}
\usepackage{booktabs}
\usepackage{pgfplots}
\pgfplotsset{compat=1.17}
\usepackage[hidelinks]{hyperref}
\usepackage{orcidlink}

\begin{document}

\title{ECGWmark: Deep Watermarking of Electrocardiogram Signals With Error-Correcting Codes}

\author{Kuan-Yu~Chen\,\orcidlink{0009-0001-7256-7263}, 
Kuan-Chen~Wang\,\orcidlink{0000-0002-7656-2822}, 
Yu~Tsao\,\orcidlink{0000-0001-6956-0418},~\IEEEmembership{Senior~Member,~IEEE}, 
and~Jian-Jiun~Ding\,\orcidlink{0000-0003-4510-2273},~\IEEEmembership{Senior~Member,~IEEE}%
\thanks{Manuscript received XXX; revised XXX; accepted XXX. (\textit{Corresponding authors: Jian-Jiun Ding and Yu Tsao.})}
\thanks{Kuan-Yu Chen, Kuan-Chen Wang, and Jian-Jiun Ding are with the Graduate Institute of Communication Engineering, National Taiwan University, Taipei 106, Taiwan (e-mail: f13942135@ntu.edu.tw; d12942016@ntu.edu.tw; jjding@ntu.edu.tw).}
\thanks{Yu Tsao is with the Research Center for Information Technology Innovation (CITI), Academia Sinica, Taipei 115, Taiwan (e-mail: yu.tsao@citi.sinica.edu.tw).}}


\maketitle

\begin{abstract}
Wearable electrocardiogram (ECG) recordings pass through multiple networks and systems before automated interpretation, making patient metadata vulnerable to loss. Digital watermarking embeds this metadata directly into the signal. However, conventional ECG watermarking methods fail under temporal desynchronization, and waveform similarity does not guarantee downstream diagnostic preservation. This article presents ECGWmark, a deep watermarking framework for single-lead ECG. A fully convolutional generator embeds the payload without signal quantization, a detector decodes the message frame by frame and verifies watermark presence, and a frozen arrhythmia classifier penalizes changes that would alter diagnostic predictions. To bridge the gap between bit accuracy and whole-message recovery, payloads are protected by extended Hamming codes, validation-selected interleaving, and soft-decision decoding. On the MIT-BIH Arrhythmia Database, ECGWmark maintains over 90\% bit accuracy under time shifts of 10 to 50 samples, where transform baselines fall to chance level. With the coding layer, the protected 16-bit payload achieves 99.86\% exact recovery on clean segments and sustains over 90\% recovery under time shifts, 10\% sample deletion, and 10\% packet loss. Inter-patient evaluation confirms improved ventricular-beat sensitivity with preserved macro-F1. Requiring only 0.56 million parameters and about 15~ms per segment on a desktop CPU, ECGWmark highlights the need to assess physiological watermarking through whole-message recovery and diagnostic consistency.
\end{abstract}

\begin{IEEEkeywords}
Arrhythmia classification, data integrity, deep watermarking, electrocardiogram (ECG), error-correcting codes, remote patient monitoring.
\end{IEEEkeywords}

\section{Introduction}
\label{sec:intro}
\IEEEPARstart{E}{lectrocardiogram} (ECG) monitoring is rapidly shifting from clinics to IoMT-connected wearables~\cite{ozcelik2025survey}. In typical remote pipelines, single-lead signals are collected on-body, transmitted over public networks, stored in health databases, and analyzed by automated models on-device or in the cloud~\cite{hannun2019cardiologist,luz2016ecg,mao2026ecgpatch,chen2023srecg}. Throughout this process, recordings risk detaching from patient identity, recording context, or source devices. While authentication protocols~\cite{soleymani2023privacy} and ledger systems~\cite{liu2026xai} secure networks and storage, they maintain metadata in separate records. Digital watermarking solves this by embedding metadata directly into the waveform, keeping it bound through format conversions to the point of interpretation~\cite{rani2026review}. Although common for medical images~\cite{bayari2026watermarking,min2026reversiblewm}, ECG signals require similar in-signal protection.

Most existing methods rely on heuristic embedding, including LSB substitution~\cite{halder2014patient}, reversible schemes~\cite{bhalerao2022reversible}, and transform-domain approaches using wavelets~\cite{ibaida2013wavelet,hamami2026securing}, SVD~\cite{jero2014dwt,anand2020improved}, curvelets~\cite{goyal2020improved}, or Schur decomposition~\cite{tiwari2026schurmark}. While these confirm that ECG signals can carry metadata with minimal distortion, three key issues hinder their practical deployment in automated pipelines:
\begin{enumerate}
    \item \textit{Failure under time shifts.} Transform-domain methods divide signals into fixed blocks, requiring the receiver to match the exact same grid. Common transmission issues such as buffer offsets, packet loss, or resampling break this alignment and cause extraction to fail~\cite{voloshynovskiy2001attacks}. In our tests, shifting just 10 samples drops bit accuracy to random chance ($\approx$50\%). Moreover, coefficient quantization often creates visible step-like artifacts in subtle regions like the P wave and ST segment.
    \item \textit{Waveform fidelity does not mean diagnostic safety.} Most studies evaluate distortion using only signal metrics like SNR~\cite{hamami2026securing}. However, deep learning classifiers can misclassify under minor, structured noise~\cite{han2020deep,finlayson2019adversarial}. Thus, a high SNR alone cannot ensure that automated diagnoses remain unchanged.
    \item \textit{High bit accuracy hides packet-level failure.} Average bit accuracy is a misleading metric because metadata like patient IDs requires zero bit errors to be usable. Assuming independent errors, a seemingly high 95\% bit accuracy yields a fully correct 32-bit message only $0.95^{32} \approx 19\%$ of the time.
\end{enumerate}

End-to-end deep watermarking can tackle these issues, but physiological signal applications remain limited~\cite{hsu2025ecg,zhang2026robust}, leaving time shifts, diagnostic safety, and full-message recovery unaddressed together. To eliminate alignment issues while preserving cardiac features, we use a lightweight, fully convolutional, quantization-free network that adds smooth residuals and decodes in real time. For diagnostic reliability, the generator trains with feedback from a downstream classifier. Finally, to ensure complete recovery, outputs are protected by error-correcting codes (ECC)~\cite{hamming1950error,lin2004errorcontrol} using soft bit probabilities.

We therefore propose \textbf{ECGWmark}, a small (0.56 million parameters), fully convolutional, and quantization-free deep watermarking framework for single-lead ECG. Our main contributions are as follows:
\begin{itemize}
    \item \textbf{Frame-level, quantization-free ECG watermarking:} A generator--detector pair embeds and reads the message with local 1-D residual convolutions on the full-resolution signal and is trained end to end through a differentiable distortion layer. This design needs no global block alignment and keeps 90.27\%--93.88\% bit accuracy under circular shifts of 10 to 50 samples.
    \item \textbf{Diagnostic fidelity constraint:} During training, a frozen arrhythmia classifier adds a differentiable penalty when the watermark changes its output. We evaluate the effect under the inter-patient protocol with macro-F1, per-class sensitivity, and an independently trained classifier, and we report the drop in supraventricular-beat sensitivity.
    \item \textbf{Error-correcting payload protection:} We show why bit accuracy and whole-message recovery differ and add a rate-0.5 coding layer with extended Hamming(8,4) codes, a channel ordering chosen on validation data, and soft maximum-likelihood decoding. This raises error-free recovery of a 16-bit message on clean segments to $99.86\pm0.08\%$.
    \item \textbf{Broad evaluation with clear scope:} We test ECGWmark under 20 distortion conditions, separating distortions seen in training from those used only for testing, and also report a comparison at equal payload and SNR, sample deletion, packet loss, watermark-presence detection, and run time.
\end{itemize}

The rest of this article is organized as follows. Section~\ref{sec:related} reviews related work. Section~\ref{sec:method} details the proposed ECGWmark framework. Section~\ref{sec:setup} outlines the experimental setup, Section~\ref{sec:results} presents the results, Section~\ref{sec:discussion} discusses the design implications and limitations, and Section~\ref{sec:conclusion} concludes the article.

\section{Related Work}
\label{sec:related}
\subsection{Conventional ECG Watermarking}
Traditional ECG watermarking relies on handcrafted rules balancing capacity, reversibility, and robustness. Time-domain LSB substitution~\cite{halder2014patient} offers high capacity but breaks under simple scaling or filtering. Reversible schemes~\cite{bhalerao2022reversible} restore exact waveforms for tamper detection, yet fail under irreversible channel noise. For better noise resistance, transform-domain methods modify wavelet sub-bands~\cite{ibaida2013wavelet,hamami2026securing}, wavelet singular values~\cite{jero2014dwt,anand2020improved}, curvelets~\cite{goyal2020improved}, or Schur matrices~\cite{tiwari2026schurmark} (reviewed in~\cite{rani2026review}).

Despite different transforms, these methods share a bottleneck: slicing the 1-D ECG into fixed blocks or 2-D matrices requires the decoder to reconstruct the exact sample grid. In practice, timing shifts from buffer offsets, packet drops, or resampling break this alignment and cause extraction failure~\cite{voloshynovskiy2001attacks}. Furthermore, vital low-amplitude features like the P wave and ST segment often suffer from step-like artifacts caused by coefficient quantization.

\subsection{Deep Learning-Based Watermarking}
End-to-end deep watermarking, introduced by HiDDeN~\cite{zhu2018hidden}, jointly trains an encoder and decoder using a differentiable distortion layer to simulate transmission channels~\cite{tancik2020stegastamp,luo2020distortion,zhang2025distortion}. While early methods spread messages across entire images~\cite{zhu2018hidden,ahmadi2020redmark,fernandez2023stable} or audio clips~\cite{chen2023wavmark,liu2024timbre}, AudioSeal~\cite{sanroman2024proactive} introduced sample-level detection to tolerate time cropping and splicing.

However, image and audio models rely on perceptual losses based on human vision or hearing. In ECGs, subtle low-amplitude waveforms which such losses might ignore as background often carry vital diagnostic cues. ECGWmark adopts the local, frame-level design of AudioSeal~\cite{sanroman2024proactive} but replaces perceptual objectives with cardiac-specific spectral and diagnostic losses. Deep watermarking for physiological signals remains rare: Hsu \textit{et al.}~\cite{hsu2025ecg} used a variational autoencoder for ECG sensor assessment, while Zhang \textit{et al.}~\cite{zhang2026robust} applied a dual-stream attention network to EEG. Neither work addresses time shifts, diagnostic safety, and full-message recovery together.

\subsection{Diagnostic Effect of Watermarks on ECG Classifiers}
Prior ECG watermarking studies typically evaluate signal quality using waveform metrics like SNR or percentage root-mean-square difference~\cite{rani2026review,hamami2026securing}. In modern clinical pipelines, however, watermarked signals are analyzed by automated 1-D convolutional or residual arrhythmia classifiers~\cite{hannun2019cardiologist,xu2019endtoend,he2016deep,kachuee2018ecg}. Two factors explain why high SNR alone cannot guarantee diagnostic safety. First, deep classifiers often rely on features different from those prioritized by cardiologists~\cite{mehari2024feature}, so a high-SNR watermark can still alter features that drive model predictions. Second, deep ECG models are vulnerable to minor, structured perturbations~\cite{han2020deep,finlayson2019adversarial,newaz2020adversarial}. Because a learned watermark is itself a structured perturbation across the signal, its diagnostic impact must be constrained during training and evaluated directly with class-sensitive metrics under an inter-patient protocol~\cite{dechazal2004automatic,luz2016ecg,mondejar2019heartbeat}, where rare arrhythmia classes are particularly sensitive.

\subsection{Error-Correcting Codes for Watermark Payloads}
A watermarking system acts as a communication channel disturbed by both the host signal and transmission noise, meaning message reliability depends on both raw bit error rates and channel coding. Extended Hamming codes provide single-error correction and double-error detection (SECDED)~\cite{hamming1950error,lin2004errorcontrol}, while soft-decision decoding further boosts performance by using bit-level confidence~\cite{chase1972class}. Although deep image watermarking has used BCH codes~\cite{tancik2020stegastamp,luo2020distortion}, these methods typically rely on hard 0/1 decisions, discarding the soft bit probabilities output by the neural detector. Similarly, prior ECG studies report average bit accuracy or correlation rather than full-message recovery rates. In Section~\ref{sec:ecc}, we pair extended Hamming codes with reliability-based channel ordering and soft maximum-likelihood decoding to exploit the uneven bit confidence of our detector.

\section{Proposed Method}
\label{sec:method}
\begin{figure*}[t]
    \centering
    \includegraphics[width=0.82\textwidth]{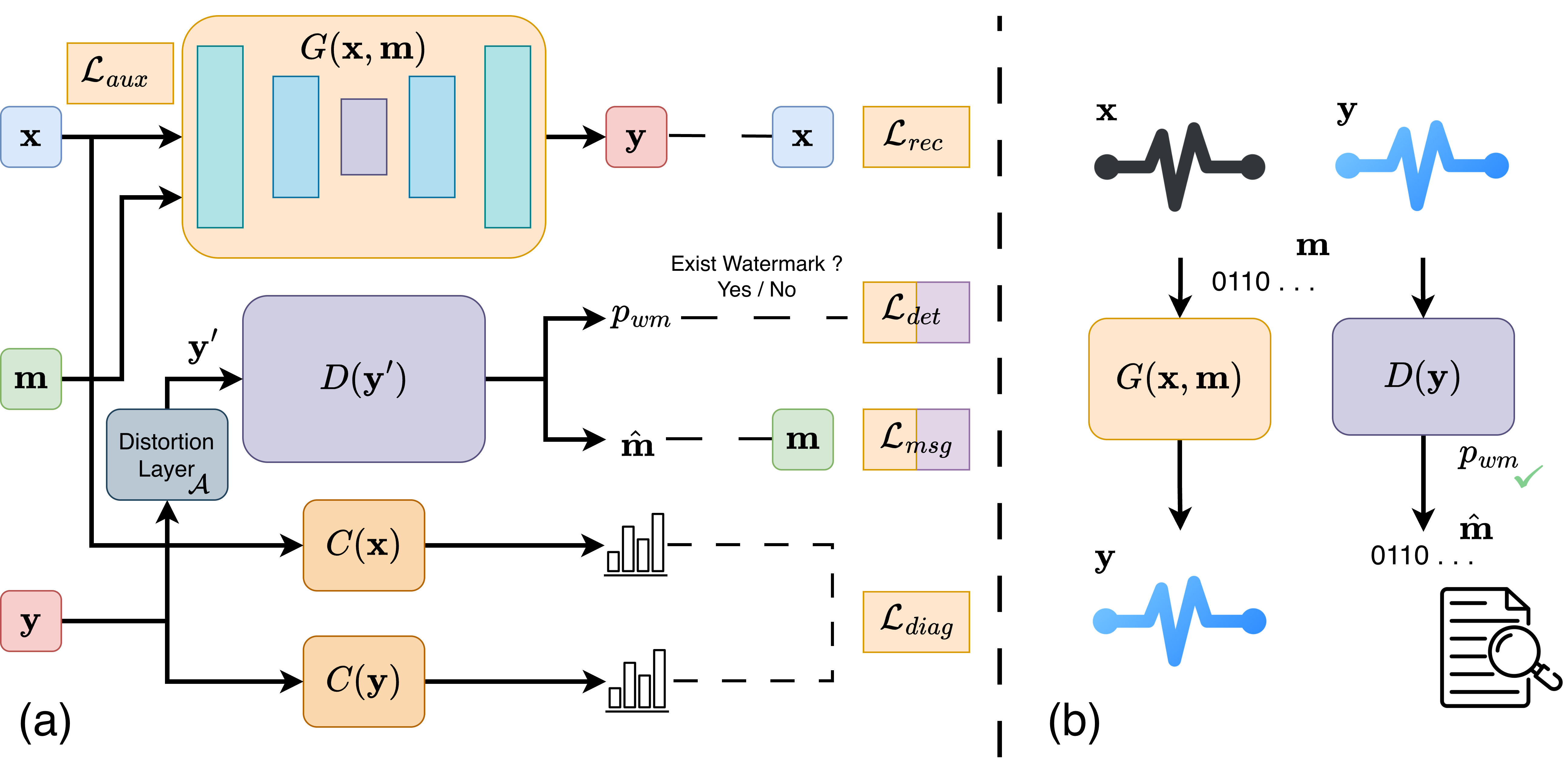}
    \caption{Overview of ECGWmark. (a) \textbf{Training.} The generator $G$ and the detector $D$ are trained together through a differentiable distortion layer $\mathcal{A}$. From an ECG segment $\mathbf{x}$ and a binary message $\mathbf{m}$, $G$ produces a watermarked ECG $\mathbf{y}$. $G$ is guided by the reconstruction loss $\mathcal{L}_{rec}$, the spectral loss $\mathcal{L}_{aux}$, and the diagnostic loss $\mathcal{L}_{diag}$, which keeps the class probabilities of a frozen classifier $C$ (bar charts) the same for $\mathbf{x}$ and $\mathbf{y}$. $D$ reads the distorted ECG $\mathbf{y}'$ and is trained with the presence loss $\mathcal{L}_{det}$ and the message loss $\mathcal{L}_{msg}$; their gradients also reach $G$ through $\mathcal{A}$. (b) \textbf{Use.} $G$ embeds $\mathbf{m}$ into $\mathbf{x}$. After transmission, $D$ estimates the presence probability $p_{wm}$ and the message $\hat{\mathbf{m}}$, and $\mathbf{y}$ can be passed to clinical analysis without realignment.}
    \label{fig:architecture}
\end{figure*}

Fig.~\ref{fig:architecture} shows the overall design of ECGWmark. The error-correcting code layer (Section~\ref{sec:ecc}) is added around the trained generator and detector and does not change their weights.

\subsection{Problem Formulation and Design Requirements}
\label{sec:problem}
Let $\mathbf{x}\in\mathbb{R}^{T}$ be a single-lead ECG segment with $T$ samples, and let $\mathbf{m}\in\{0,1\}^{N}$ be the binary message, where $N$ is the number of \emph{channel bits} per segment. The generator $G$ embeds the message by adding a scaled, learned residual $R_{\theta}(\mathbf{x},\mathbf{m})$ to the signal:
\begin{equation}
\mathbf{y}=G(\mathbf{x},\mathbf{m})=\mathbf{x}+\alpha\,R_{\theta}(\mathbf{x},\mathbf{m}),
\label{eq:embed}
\end{equation}
where $\mathbf{y}\in\mathbb{R}^{T}$ is the watermarked segment and $\alpha$ is a global embedding strength. During transmission and storage, the signal goes through an unknown distortion $\mathcal{T}$, and the receiver gets $\mathbf{y}'=\mathcal{T}(\mathbf{y})$. The detector $D$ then reads $\mathbf{y}'$ directly:
\begin{equation}
(p_{wm},\mathbf{q})=D(\mathbf{y}'),
\label{eq:detect}
\end{equation}
and outputs the probability $p_{wm}\in[0,1]$ that a watermark is present and the bit probabilities $\mathbf{q}\in[0,1]^{N}$. Without coding, each bit is decided by a threshold, $\hat{m}_i=\mathbf{1}[q_i>0.5]$. In this study, each segment has $T=1{,}024$ samples ($\approx2.84$~s at 360~Hz) and carries $N=32$ channel bits.

Extraction is \emph{blind}: the detector needs neither the original ECG $\mathbf{x}$ nor any extra synchronization information. We consider non-malicious distortions $\mathcal{T}$ that commonly occur during acquisition, transmission, and preprocessing (Section~\ref{sec:attacks}). ECGWmark does not use a secret key or a digital signature, so protection against deliberate removal or forgery is outside the scope of this work. The design has three goals: \emph{waveform fidelity} ($\mathbf{y}\approx\mathbf{x}$), \emph{diagnostic consistency} (an arrhythmia classifier $C$ gives the same output for $\mathbf{x}$ and $\mathbf{y}$), and \emph{message reliability} (the whole message is recovered correctly from $\mathbf{y}'$).

\subsection{Frame-Level Neural Architecture}
\label{sec:arch}
Both $G$ and $D$ are fully convolutional networks that work on the 1-D ECG at its original sampling rate. The generator has an encoder--bottleneck--decoder structure made of 1-D residual convolution blocks. The message $\mathbf{m}$ is stretched along the time axis and merged with the ECG features in the bottleneck through residual convolutions. The features stay continuous: no vector quantization or codebook lookup is used. This lets the residual $R_{\theta}$ make smooth, small changes to the waveform instead of the step-like changes that a discrete bottleneck can cause.

The detector $D$ uses local 1-D convolutions to estimate watermark presence and message bits at every time step (frame). These frame-level estimates are then averaged over time to give $p_{wm}$ and $\mathbf{q}$ for the whole segment. Because convolution is shift-equivariant, a time shift of the input mainly moves the frame-level evidence instead of destroying it, so the averaged output depends little on exact sample alignment. Unlike block-transform methods, the detector does not need to rebuild a fixed coefficient grid. This makes it robust to circular shifts and lets it recover part of the evidence when some samples are deleted; when too much of the signal is lost, however, performance still drops (Section~\ref{sec:res_ecc}).

\subsection{Differentiable Distortion Layer}
\label{sec:distortion}
To make the watermark robust, a random distortion layer $\mathcal{A}$ is placed between $G$ and $D$ during training. For each mini-batch, $\mathcal{A}$ applies, with a preset probability, additive white Gaussian noise (AWGN), 1-D Gaussian smoothing, amplitude scaling, or a circular time shift. All four operations are differentiable, so the gradients from $D$ flow back through $\mathcal{A}$ into $G$. This pushes the generator to place the watermark in signal components that survive these distortions rather than in fragile high-frequency details. Like the noise layers used in image watermarking~\cite{zhu2018hidden,tancik2020stegastamp,zhang2025distortion}, $\mathcal{A}$ simulates the expected transmission channel. To test distortions not seen in training, band-pass filtering, amplitude quantization, resampling, sample deletion, and packet loss are used only for evaluation (Section~\ref{sec:attacks}).

\subsection{Training Objectives}
\label{sec:losses}
The generator and detector are trained together by minimizing
\begin{equation}
\begin{aligned}
\mathcal{L}_{total}={}&\lambda_{rec}\mathcal{L}_{rec}+\lambda_{msg}\mathcal{L}_{msg}+\lambda_{det}\mathcal{L}_{det}\\
&+\lambda_{diag}\mathcal{L}_{diag}+\mathcal{L}_{aux},
\end{aligned}
\label{eq:total}
\end{equation}
where the weights $\lambda$ balance waveform fidelity, message reliability, and diagnostic consistency.

\noindent\textbf{Reconstruction and spectral losses.} The mean squared error (MSE) keeps the watermarked ECG $\mathbf{y}$ close to the original $\mathbf{x}$ in the time domain:
\begin{equation}
\mathcal{L}_{rec}=\frac{1}{T}\sum_{t=1}^{T}(x_t-y_t)^2 .
\end{equation}
The MSE alone does not control which frequency bands carry the watermark, so we add spectral losses in the Fourier and continuous wavelet domains~\cite{schwar2023multi}:
\begin{equation}
\mathcal{L}_{aux}=\lambda_{fft}\big\|\mathcal{F}(\mathbf{x})-\mathcal{F}(\mathbf{y})\big\|_2^2+\lambda_{cwt}\big\|\mathcal{W}(\mathbf{x})-\mathcal{W}(\mathbf{y})\big\|_2^2,
\end{equation}
where $\mathcal{F}(\cdot)$ and $\mathcal{W}(\cdot)$ denote the fast Fourier transform and continuous wavelet transform, respectively.

\noindent\textbf{Message and detection losses.} Message decoding is trained with the binary cross-entropy (BCE) between the embedded bits $\mathbf{m}$ and the predicted probabilities $\mathbf{q}$:
\begin{equation}
\mathcal{L}_{msg}=-\frac{1}{N}\sum_{i=1}^{N}\Big[m_i\log q_i+(1-m_i)\log(1-q_i)\Big].
\end{equation}
To avoid false detections on ECGs without a watermark, distorted clean segments $\mathbf{x}'\sim\mathcal{A}(\mathbf{x})$ are also given to $D$ during training, with a binary presence loss~\cite{sanroman2024proactive}:
\begin{equation}
\begin{aligned}
\mathcal{L}_{det}=-\Big[&\mathbb{E}_{\mathbf{y}'\sim\mathcal{A}(\mathbf{y})}\log p_{wm}(\mathbf{y}')\\
&+\mathbb{E}_{\mathbf{x}'\sim\mathcal{A}(\mathbf{x})}\log\big(1-p_{wm}(\mathbf{x}')\big)\Big].
\end{aligned}
\end{equation}

\noindent\textbf{Diagnostic fidelity loss.} A high SNR does not guarantee that a classifier gives the same diagnosis for $\mathbf{x}$ and $\mathbf{y}$~\cite{han2020deep}. We therefore use a frozen, pre-trained five-class arrhythmia classifier $C$ based on a 1-D ResNet~\cite{he2016deep,kachuee2018ecg} and penalize the Kullback--Leibler (KL) divergence between its outputs:
\begin{equation}
\mathcal{L}_{diag}=D_{\mathrm{KL}}\big(C(\mathbf{x})\,\|\,C(\mathbf{y})\big)=\sum_{k=1}^{K}C_k(\mathbf{x})\log\frac{C_k(\mathbf{x})}{C_k(\mathbf{y})},
\end{equation}
where $K=5$ is the number of AAMI heartbeat classes. Gradients pass through the frozen $C$ into $G$ and discourage residuals that change the prediction. Because only one classifier guides training, we also check diagnostic consistency with class-level metrics and a separately trained classifier (Section~\ref{sec:res_diag}).

\subsection{Training Dynamics}
\label{sec:training}
Three simple measures keep the joint training stable. First, the embedding strength $\alpha$ in \eqref{eq:embed} is lowered from 0.5 to 0.15 with a cosine schedule~\cite{loshchilov2016sgdr}, so the detector receives a strong signal early on and the watermark becomes less visible later. Second, following the two-time-scale update rule~\cite{heusel2017gans}, the learning rate of $D$ is $1.5\times$ that of $G$. Third, an exponential moving average (decay 0.999) of the weights of both networks is kept to reduce fluctuations on the validation set.

\subsection{Payload Protection With Error-Correcting Codes}
\label{sec:ecc}
The trained network gives $N=32$ noisy bit channels per segment. We first explain why sending 32 uncoded bits over these channels gives poor message reliability, and then describe a light coding layer that turns them into a reliable 16-bit message without retraining $G$ or $D$.

\subsubsection{From Bit Accuracy to Message Reliability}
Let $p_j$ be the error probability of channel $j$ and $\bar{p}=\frac{1}{N}\sum_{j=1}^{N} p_j$ the mean error rate, so the average bit accuracy is $1-\bar{p}$. If errors are independent, the probability that an uncoded $N$-bit message has no error is
\begin{equation}
P_{\mathrm{exact}}^{\mathrm{unc}}=\prod_{j=1}^{N}(1-p_j)\;\le\;(1-\bar{p})^{N},
\label{eq:exact_uncoded}
\end{equation}
where the inequality follows from the inequality of arithmetic and geometric means. Equation~\eqref{eq:exact_uncoded} shows two things. First, the chance of a fully correct message falls exponentially with the message length $N$: at our measured clean bit accuracy of $1-\bar{p}=95.11\%$, even the upper bound is only $20.1\%$ for $N=32$. Second, for the same average accuracy, uneven channel reliability \emph{lowers} this chance even more. The same unevenness, however, can be used to our advantage by putting the information bits on the most reliable channels, which we do and verify in Section~\ref{sec:res_ecc}.

\subsubsection{Extended Hamming(8,4) Code}
A 16-bit message $\mathbf{u}\in\{0,1\}^{16}$ is split into four 4-bit blocks $\mathbf{u}^{(b)}$ ($b=1,\ldots,4$). Each block is encoded with the systematic extended Hamming(8,4) code $\mathcal{C}_{8,4}$~\cite{hamming1950error,lin2004errorcontrol} as $\mathbf{c}^{(b)}=\mathbf{u}^{(b)}\mathbf{G}_{8,4}$ over $\mathrm{GF}(2)$, with the generator matrix
\begin{equation}
\mathbf{G}_{8,4}=\left[\begin{array}{cccc|cccc}
1&0&0&0&1&1&0&1\\
0&1&0&0&1&0&1&1\\
0&0&1&0&0&1&1&1\\
0&0&0&1&1&1&1&0
\end{array}\right],
\label{eq:generator}
\end{equation}
where the first four columns copy the information bits and the last four add three Hamming parity bits and one overall parity bit (other systematic forms only reorder the parity columns). Each block has 16 valid codewords and a minimum Hamming distance of $d_{\min}=4$, so hard decoding corrects one error and detects two (SECDED). The four 8-bit codewords together form the $N=32$ channel bits embedded by $G$, giving a code rate of $R=0.5$, i.e., 16 information bits per segment.

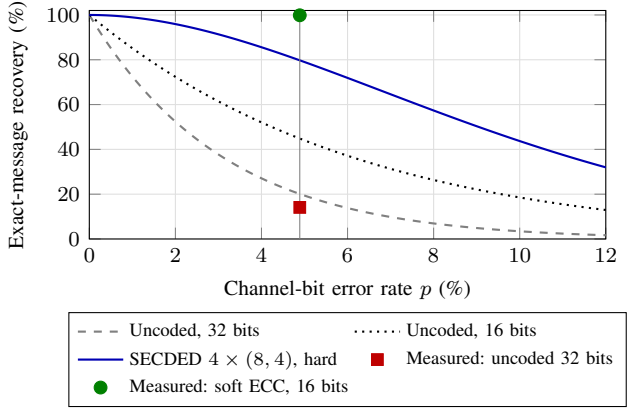
\begin{figure}[t]
\centering
\begin{tikzpicture}
\begin{axis}[
    width=0.95\linewidth, height=4.6cm,
    xlabel={Channel-bit error rate $p$ (\%)},
    ylabel={Exact-message recovery (\%)},
    xmin=0, xmax=12, ymin=0, ymax=102,
    grid=major, grid style={gray!25},
    legend style={font=\scriptsize, at={(0.5,-0.32)}, anchor=north, legend columns=2, cells={anchor=west}},
    tick label style={font=\footnotesize}, label style={font=\footnotesize},
]
\addplot[thick, dashed, color=gray, domain=0:12, samples=121] {100*(1-x/100)^32};
\addlegendentry{Uncoded, 32 bits}
\addplot[thick, dotted, color=black, domain=0:12, samples=121] {100*(1-x/100)^16};
\addlegendentry{Uncoded, 16 bits}
\addplot[thick, color=blue!70!black, domain=0:12, samples=121] {100*((1-x/100)^8 + 8*(x/100)*(1-x/100)^7)^4};
\addlegendentry{SECDED $4\times(8,4)$, hard}
\addplot[only marks, mark=square*, mark size=2.2pt, color=red!75!black] coordinates {(4.89,14.10)};
\addlegendentry{Measured: uncoded 32 bits}
\addplot[only marks, mark=*, mark size=2.4pt, color=green!50!black] coordinates {(4.89,99.86)};
\addlegendentry{Measured: soft ECC, 16 bits}
\draw[gray, thin] (axis cs:4.89,0) -- (axis cs:4.89,102);
\end{axis}
\end{tikzpicture}
\caption{Exact-message recovery versus channel-bit error rate $p$ when all channels have the same independent error rate (curves, from \eqref{eq:exact_uncoded} and \eqref{eq:exact_hard}). Markers are measured on clean DS2 segments at the observed mean error rate $p=4.89\%$ (95.11\% bit accuracy). The uncoded 32-bit result lies below its curve because channel reliability is uneven; the soft ECC result lies far above the hard-decoding curve because it combines reliability-based channel ordering with soft ML decoding.}
\label{fig:ecc_theory}
\end{figure}

If all channels have the same independent error rate $p$, hard decoding of a block succeeds when it contains at most one error:
\begin{equation}
P_{\mathrm{blk}}^{\mathrm{hard}}=(1-p)^{8}+8p(1-p)^{7},\qquad P_{\mathrm{exact}}^{\mathrm{hard}}=\big(P_{\mathrm{blk}}^{\mathrm{hard}}\big)^{4}.
\label{eq:exact_hard}
\end{equation}
At the measured clean error rate $p=4.89\%$, \eqref{eq:exact_hard} predicts $79.8\%$ fully correct messages, compared with $20.1\%$ for uncoded 32 bits and $44.8\%$ for uncoded 16 bits (Fig.~\ref{fig:ecc_theory}). Our full coding layer reaches a much higher rate ($99.86\%$) because it adds two steps that use the uneven channel reliability and the bit probabilities.

\subsubsection{Reliability-Based Channel Ordering}
A fixed permutation $\pi$ maps the 32 codeword positions to the 32 channels of the network. To build $\pi$, we measure the error rate $\hat{\varepsilon}_j$ of each channel $j$ \emph{only on a held-out validation subset of DS1}, averaged over the clean condition and circular shifts of $+10$, $+30$, and $+50$ samples. The 16 most reliable channels (lowest $\hat{\varepsilon}_j$) carry the information bits, and the other 16 carry the parity bits. $\pi$ is fixed before any DS2 testing and kept the same for all seeds and distortions. At the receiver, the probabilities are reordered back as $\tilde{\mathbf{q}}=\pi^{-1}(\mathbf{q})$ before decoding.

\subsubsection{Soft-Decision Maximum-Likelihood Decoding}
A hard threshold at $0.5$ throws away the confidence information in $\mathbf{q}$. Following soft-decision decoding~\cite{chase1972class}, we decode each 8-bit block by exact maximum likelihood (ML) over all 16 valid codewords. Let $\tilde{\mathbf{q}}^{(b)}\in[0,1]^8$ be the reordered bit probabilities of block $b$. Treating the channels as independent Bernoulli variables gives the soft ML rule
\begin{equation}
\hat{\mathbf{c}}^{(b)}=\underset{\mathbf{c}\in\mathcal{C}_{8,4}}{\arg\max}\sum_{i=1}^{8}\Big[c_i\log\tilde{q}^{(b)}_i+(1-c_i)\log\big(1-\tilde{q}^{(b)}_i\big)\Big].
\label{eq:softml}
\end{equation}
Removing terms that do not depend on $\mathbf{c}$, \eqref{eq:softml} becomes the maximization of $\sum_{i=1}^{8} c_i\ell^{(b)}_i$ with the log-likelihood ratios $\ell^{(b)}_i=\log\big(\tilde{q}^{(b)}_i/(1-\tilde{q}^{(b)}_i)\big)$; probabilities are clipped to $[\epsilon,1-\epsilon]$ for numerical stability. The four information bits of each decoded codeword $\hat{\mathbf{c}}^{(b)}$ form the recovered 16-bit message $\hat{\mathbf{u}}$. Hard decoding fails when two or more bits in a block are wrong. Soft ML decoding can still succeed in that case if the wrong bits have low confidence and the information bits have high confidence, which is what the reliability-based channel ordering aims for.

Checking all $4\times16=64$ codewords per segment needs only simple inner products, which is negligible compared with one pass through the network. The coding layer has no trainable parameters, so the code can be changed after training. Because ML decoding always returns the closest valid codeword, it corrects errors but does not reject badly damaged segments unless a confidence threshold is added, and it does not replace cryptographic authentication.

\section{Experimental Setup}
\label{sec:setup}
\subsection{Dataset and Evaluation Protocol}
\label{sec:dataset}
All experiments use the MIT-BIH Arrhythmia Database \cite{moody2001impact}, which contains 48 half-hour, two-channel ambulatory recordings sampled at 360~Hz with cardiologist-verified beat labels. The dataset is publicly available via PhysioNet~\cite{goldberger2000physiobank}\footnote{\url{https://physionet.org/content/mitdb/1.0.0/}}, and because it is public and de-identified, no ethical approval was required. All segments are single-lead and $T=1{,}024$ samples long ($\approx2.84$~s). Following the inter-patient protocol of de Chazal \textit{et al.}~\cite{dechazal2004automatic}, we exclude the four paced records (102, 104, 107, and 217) and split the remaining 44 records equally into DS1 (records 101, 106, 108, 109, 112, 114, 115, 116, 118, 119, 122, 124, 201, 203, 205, 207, 208, 209, 215, 220, 223, and 230) and DS2 (records 100, 103, 105, 111, 113, 117, 121, 123, 200, 202, 210, 212, 213, 214, 219, 221, 222, 228, 231, 232, 233, and 234), with 22 records each. After preprocessing, DS1 contains 50,999 labeled beats and DS2 contains 49,690. Message-level evaluation uses 49,634 beat-centered DS2 segments, with beat labels mapped to the five AAMI classes (N, S, V, F, and Q).

We partition the data as follows: (i) the diagnostic classifier $C$ is trained on DS1 and tested on DS2; (ii) the watermarking network is trained on randomly split segments rather than a patient-separated split (see Section~\ref{sec:limitations} for implications); and (iii) all tunable parameters namely baseline embedding strengths (Section~\ref{sec:baselines}) and ECC channel ordering (Section~\ref{sec:ecc}) are selected on a held-out DS1 validation subset and fixed prior to DS2 testing.

\subsection{Baselines}
\label{sec:baselines}
We compare ECGWmark against three representative handcrafted methods: LSB substitution~\cite{halder2014patient}, a DCT-SVD method representing SVD-based transform watermarking~\cite{jero2014dwt,anand2020improved}, and SchurMark~\cite{tiwari2026schurmark}. We evaluate two settings:
\begin{itemize}
    \item \textbf{Original setting} (Table~\ref{tab:attacks_comparison}). Each baseline embeds its default 16-bit payload, whereas ECGWmark embeds 32 channel bits. Since payload size and distortion levels differ, this setting simply illustrates how each method responds to individual distortions.
    \item \textbf{Matched setting} (Table~\ref{tab:matched}). All methods embed 16 information bits, with ECGWmark using the rate-0.5 ECC layer (ECGWmark-ECC16). The embedding strengths of DCT-SVD and SchurMark are tuned on the DS1 validation subset to match the clean SNR of ECGWmark-ECC16, then fixed. Notably, both transform baselines require the original signal for extraction (non-blind), whereas ECGWmark operates blindly.
\end{itemize}

\subsection{Distortions}
\label{sec:attacks}
We evaluate robustness across 20 distortion conditions. During training, the distortion layer $\mathcal{A}$ incorporates four categories: AWGN (10, 15, 20, and 30~dB SNR), Gaussian smoothing ($\sigma=1$ and $3$ samples), amplitude scaling ($0.8\times$ and $1.2\times$), and circular time shifts ($+10$, $+30$, and $+50$ samples). To test generalization to unseen channel impairments, we further introduce band-pass filtering (0.5--100~Hz), 8-bit and 6-bit amplitude quantization, resampling by factors of 2 and 4, 10\% sample deletion at the front or middle of the segment, and packet loss of 10\% and 50\%. We distinguish between desynchronization without data loss (circular shifts) and true data-loss conditions (sample deletion and packet loss), reporting their results separately for clarity.

\subsection{Metrics}
\label{sec:metrics}
\noindent\textbf{Fidelity.} The SNR of the watermarked signal is $\mathrm{SNR}=10\log_{10}\big(\|\mathbf{x}\|_2^2/\|\mathbf{x}-\mathbf{y}\|_2^2\big)$.

\noindent\textbf{Payload.} \emph{Channel-bit accuracy} is the share of channel bits that are decoded correctly. \emph{Exact-message recovery} is the share of segments in which \emph{all} message bits are correct: 32 bits for the uncoded message and 16 bits for the coded message. For the ECC analysis, each of three random seeds draws new messages and distortions for all 49,634 DS2 segments, and we report the mean and standard deviation over seeds.

\noindent\textbf{Detection.} Watermark-presence detection is measured on clean and watermarked DS2 segments with the area under the ROC curve (AUROC) and the false-positive (FPR) and false-negative (FNR) rates at the threshold $p_{wm}=0.5$.

\noindent\textbf{Diagnosis.} Most DS2 beats are Class N, so overall accuracy alone is misleading; we also report macro-F1 and the sensitivity of each class (N, S, V, F, Q)~\cite{luz2016ecg}. Unless stated otherwise, the classifier is trained on clean DS1 and applied, without retraining, to watermarked DS2 signals (Clean$\rightarrow$Watermarked). This matches a deployment in which an already-validated model is kept fixed.

\subsection{Implementation}
\label{sec:impl}
The payload has $N=32$ channel bits, the final embedding strength is $\alpha=0.15$, and training follows Section~\ref{sec:training}. Run time is measured with batch size one on an x86\_64 desktop CPU and on an NVIDIA RTX~4090 GPU.

\section{Results}
\label{sec:results}
\begin{table}[t]
\centering
\caption{Channel-Bit Accuracy (\%) Under Distortions in the Original (Unmatched) Setting.}
\label{tab:attacks_comparison}
\footnotesize
\setlength{\tabcolsep}{2.2pt}
\renewcommand{\arraystretch}{1.08}
\begin{tabular}{@{}l l c c c c@{}}
\toprule
\multirow{2}{*}{\textbf{Group}} & \multirow{2}{*}{\textbf{Condition}} & \multicolumn{4}{c}{\textbf{Method}} \\
\cmidrule(l){3-6}
 & & \textbf{ECGWmark} & \textbf{LSB} & \textbf{DCT-SVD} & \textbf{Schur.} \\
 & & \textbf{(Ours)} & \cite{halder2014patient} & \cite{jero2014dwt,anand2020improved} & \cite{tiwari2026schurmark} \\
 & Payload (bits) & 32 & 16 & 16 & 16 \\
\midrule
\multirow{3}{*}{Shift$^{\dagger}$} & $+10$ & \textbf{93.88} & 50.08 & \underline{50.91} & 50.86 \\
 & $+30$ & \textbf{90.27} & \underline{51.14} & 50.55 & 50.47 \\
 & $+50$ & \textbf{91.19} & 50.15 & \underline{50.77} & 50.10 \\
\midrule
\multirow{2}{*}{Resamp.} & $\downarrow2\uparrow2$ & \textbf{95.04} & 50.12 & 68.64 & \underline{77.79} \\
 & $\downarrow4\uparrow4$ & \textbf{91.26} & 50.17 & \underline{62.53} & 55.60 \\
\midrule
Filter & BP 0.5--100 Hz & \textbf{95.02} & 49.66 & 63.60 & \underline{67.83} \\
\midrule
\multirow{2}{*}{Smooth.$^{\dagger}$} & $\sigma=1$ & \textbf{94.99} & 50.03 & 60.09 & \underline{62.09} \\
 & $\sigma=3$ & \textbf{92.31} & 51.18 & 50.78 & \underline{52.36} \\
\midrule
\multirow{2}{*}{Ampl.$^{\dagger}$} & Scale $\times0.8$ & \textbf{94.89} & 49.18 & 90.11 & \underline{92.95} \\
 & Scale $\times1.2$ & \textbf{95.14} & 49.18 & 91.62 & \underline{94.09} \\
\midrule
\multirow{4}{*}{AWGN$^{\dagger}$} & 10 dB & \textbf{81.44} & 53.22 & \underline{79.03} & 66.67 \\
 & 15 dB & \underline{91.01} & 54.73 & \textbf{92.31} & 76.68 \\
 & 20 dB & \underline{94.23} & 50.46 & \textbf{99.57} & 90.44 \\
 & 30 dB & \underline{95.09} & 44.11 & \textbf{100.00} & \textbf{100.00} \\
\midrule
\multirow{2}{*}{Quant.} & 8-bit & 94.95 & 50.16 & \underline{99.74} & \textbf{99.93} \\
 & 6-bit & 94.39 & 49.69 & \textbf{99.67} & \underline{99.65} \\
\midrule
None & Clean & \underline{95.27} & \textbf{100.00} & \textbf{100.00} & \textbf{100.00} \\
\midrule
\multicolumn{2}{@{}l}{\textbf{Mean (16 distorted)}} & \textbf{92.82} & 50.20 & \underline{75.62} & 74.22 \\
\multicolumn{2}{@{}l}{\textbf{Mean (all 17 cond.)}} & \textbf{92.96} & 53.13 & \underline{77.05} & 75.74 \\
\bottomrule
\multicolumn{6}{@{}p{0.97\linewidth}@{}}{\scriptsize Bold and underlined values are the best and second-best results. $^{\dagger}$Distortion type also used during ECGWmark training (Section~\ref{sec:attacks}). Shifts are in samples; Schur.\ = SchurMark. Payload lengths differ, so this table describes robustness rather than a matched comparison (see Table~\ref{tab:matched}). Bit accuracy does not tell whether the whole message is correct (see Table~\ref{tab:ecc_feasibility}).}
\end{tabular}
\end{table}

\subsection{Robustness to Signal Distortions}
\label{sec:res_robust}
Table~\ref{tab:attacks_comparison} summarizes bit accuracy across clean signals and 16 distortion conditions. Under \emph{circular shifts} ($+10$, $+30$, and $+50$ samples), ECGWmark maintains 90.27\%--93.88\% accuracy, whereas all three handcrafted baselines drop to random chance (50.08\%--51.14\%), highlighting the main strength of our frame-level design. Under \emph{resampling, band-pass filtering, and Gaussian smoothing}, ECGWmark achieves 91.26\%--95.04\% even though resampling and filtering were unseen during training while DCT-SVD and SchurMark fall to 50.78\%--77.79\%. ECGWmark also leads under \emph{amplitude scaling} (94.89\%--95.14\%) and severe \emph{10-dB AWGN} (81.44\%). Under \emph{moderate noise (15--30~dB) and amplitude quantization}, transform-domain methods score higher (e.g., 99.57\% for DCT-SVD vs.\ 94.23\% for ECGWmark at 20~dB) because spreading bits across coefficient blocks helps average out broadband noise when alignment holds. Even so, ECGWmark consistently exceeds 91\% in these settings, while LSB fails across all distortions. Overall, ECGWmark averages 92.82\% accuracy across the 16 distortions, outperforming DCT-SVD (75.62\%) and SchurMark (74.22\%) by over 17 percentage points (pp). On clean signals, ECGWmark reaches 95.27\% compared with 100\% for handcrafted methods; eliminating this residual error is the primary motivation for our error-correcting code (ECC) layer.

\subsection{Message-Level Reliability With Error-Correcting Codes}
\label{sec:res_ecc}
Across the 49,634 DS2 test segments, the network achieves a 95.11\% mean clean bit accuracy (varying slightly from the 95.27\% in Table~\ref{tab:attacks_comparison} due to a separate evaluation run). Despite this high bit-level score, an \emph{uncoded} 32-bit message is recovered without any bit errors in only 14.10\% of segments because even a single flipped bit corrupts the entire payload. This result falls below the 20.1\% theoretical bound predicted by \eqref{eq:exact_uncoded} under uniform error assumptions (Fig.~\ref{fig:ecc_theory}), showing that error rates vary unevenly across the 32 channels. Consequently, raw network outputs cannot be used directly as 32 reliable information bits without error-correcting codes.

\begin{table}[t]
\centering
\caption{Exact 16-Bit Message Recovery (\%, Mean $\pm$ SD Over Three Seeds) on 49,634 DS2 Segments.}
\label{tab:ecc_feasibility}
\footnotesize
\setlength{\tabcolsep}{5pt}
\renewcommand{\arraystretch}{1.12}
\begin{tabular}{@{}l c c c@{}}
\toprule
\textbf{Condition} & \textbf{Raw systematic} & \textbf{Soft ECC (Ours)} & \textbf{$\Delta$ (pp)} \\
\midrule
Clean & $97.33\pm0.23$ & $\mathbf{99.86\pm0.08}$ & $+2.53$ \\
Shift $+10$ & $96.29\pm0.45$ & $\mathbf{99.59\pm0.04}$ & $+3.30$ \\
Shift $+30$ & $\mathbf{92.53\pm0.62}$ & $90.41\pm0.18$ & $-2.12$ \\
Shift $+50$ & $94.66\pm0.05$ & $\mathbf{96.51\pm0.19}$ & $+1.85$ \\
Front deletion 10\% & $90.41\pm0.14$ & $\mathbf{95.06\pm0.55}$ & $+4.65$ \\
Middle deletion 10\% & $90.32\pm0.97$ & $\mathbf{98.46\pm0.31}$ & $+8.14$ \\
Packet loss 10\% & $80.47\pm0.06$ & $\mathbf{97.35\pm0.30}$ & $+16.88$ \\
Packet loss 50\% & $17.91\pm1.94$ & $\mathbf{45.54\pm2.46}$ & $+27.63$ \\
\bottomrule
\multicolumn{4}{@{}p{0.95\linewidth}@{}}{\scriptsize 16 logical bits in 32 channel bits (code rate 0.5). \emph{Raw systematic}: the 16 information channels are read directly and the parity bits are ignored. \emph{Soft ECC}: ML decoding in \eqref{eq:softml}. The channel ordering is chosen on DS1 validation data and is the same for all rows.}
\end{tabular}
\end{table}

Table~\ref{tab:ecc_feasibility} shows how often the coded 16-bit message is fully correct. To separate the effect of decoding from the effect of channel ordering, both columns use the same ordering $\pi$: \emph{raw systematic} only reads the 16 reliable information channels and ignores the parity bits, while \emph{soft ECC} uses the full ML decoder in \eqref{eq:softml}.

\noindent\textbf{Effect of channel ordering.} Reading only the information channels already gives 97.33\% fully correct messages on clean data, far above the 44.8\% expected for 16 channels with equal error rates of $p=4.89\%$. Assuming independent errors, this means that the 16 information channels have a mean error rate of at most $1-0.9733^{1/16}\approx0.17\%$. Since the mean over all 32 channels is 4.89\%, the 16 parity channels must then have a mean error rate of at least 9.6\%. The ordering thus moves the noisy channels to parity positions, where soft decoding can handle their errors.

\noindent\textbf{Effect of soft decoding.} Soft ECC is better than raw systematic decoding in seven of the eight conditions. On clean segments it reaches $99.86\pm0.08\%$, which cuts the message failure rate from 2.67\% to 0.14\%, about 19 times lower. The gain is larger when part of the signal is lost: $+8.14$~pp under 10\% middle deletion, $+16.88$~pp under 10\% packet loss, and $+27.63$~pp under 50\% packet loss, where the parity bits supply the missing information. The $+30$-sample shift is the only exception (90.41\% vs.\ 92.53\%); one possible reason is that this shift changes which channels are reliable compared with the validation data. The benefit of coding therefore depends on the distortion.

\noindent\textbf{Deletion and packet loss.} With soft ECC, the 16-bit message is fully correct in 95.06\%, 98.46\%, and 97.35\% of segments under 10\% front deletion, 10\% middle deletion, and 10\% packet loss, even though none of these distortions was used in training. Under 50\% packet loss, the rate drops to 45.54\%, which marks a clear limit (Section~\ref{sec:limitations}).

\subsection{Capacity- and SNR-Matched Comparison}
\label{sec:res_matched}
\begin{table}[t]
\centering
\caption{Matched Comparison at 16 Effective Bits: Exact Recovery (\%) on DS2.}
\label{tab:matched}
\footnotesize
\setlength{\tabcolsep}{3.5pt}
\renewcommand{\arraystretch}{1.12}
\begin{tabular}{@{}l c c c@{}}
\toprule
\multirow{2}{*}{\textbf{Condition / Property}} & \multicolumn{3}{c}{\textbf{Method}} \\
\cmidrule(l){2-4}
 & \textbf{ECGWmark-ECC16} & \textbf{DCT-SVD} & \textbf{SchurMark} \\
 & \textbf{(Ours)} & \cite{jero2014dwt,anand2020improved} & \cite{tiwari2026schurmark} \\
\midrule
Extraction & Blind & Non-blind & Non-blind \\
Clean SNR (dB) & 12.25 & 11.88 & 11.88 \\
\midrule
Shift $+10$ & \textbf{98.73} & \underline{0.00} & \underline{0.00} \\
Shift $+30$ & \textbf{75.64} & \underline{0.00} & \underline{0.00} \\
Shift $+50$ & \textbf{84.45} & \underline{0.00} & \underline{0.00} \\
AWGN 20 dB & \underline{98.05} & \textbf{100.00} & \textbf{100.00} \\
\bottomrule
\multicolumn{4}{@{}p{0.95\linewidth}@{}}{\scriptsize Bold and underlined values are the best and second-best results. Baseline strengths are tuned on DS1 validation data to match the clean SNR of ECGWmark-ECC16 and fixed before testing on DS2. ECGWmark-ECC16 also achieves 99.41\% clean and 99.23\% band-pass-filtered exact recovery; LSB~\cite{halder2014patient} is omitted because it fails under every distortion in Table~\ref{tab:attacks_comparison}. SNR values are measured within this run and therefore differ slightly from Table~\ref{tab:fidelity_comparison}(b).}
\end{tabular}
\end{table}

Table~\ref{tab:matched} removes the payload mismatch of Table~\ref{tab:attacks_comparison}: all methods carry 16 information bits at about the same clean SNR ($\approx12$~dB). Under circular shifts of $+10$, $+30$, and $+50$ samples, ECGWmark-ECC16 recovers the full message in 98.73\%, 75.64\%, and 84.45\% of segments, while DCT-SVD and SchurMark recover none. Under 20-dB AWGN, the two non-blind baselines reach 100.00\%, compared with 98.05\% for our blind detector. The comparison therefore shows a trade-off between robustness to noise and robustness to time shifts, not a method that wins everywhere. Tables~\ref{tab:ecc_feasibility} and~\ref{tab:matched} come from separate evaluation runs (with different segment subsets and random messages), so their numbers are not directly comparable: Table~\ref{tab:matched} compares the methods under the same conditions within one run, while Table~\ref{tab:ecc_feasibility} describes the coding layer over three seeds on all 49,634 segments.

\subsection{Watermark-Presence Detection}
\label{sec:res_det}
When separating clean from watermarked DS2 segments, the detector $D$ reaches an AUROC of 1.0000 (rounded to four decimals), with an FPR of 0.018\% and an FNR of 0.000\% at the threshold $p_{wm}=0.5$. The detector thus reliably tells whether a watermark is present under normal processing; because no secret key is used, this says nothing about security against deliberate forgery or removal.

\begin{table*}[t]
\centering
\caption{Diagnostic Fidelity on DS2 (\%). (a) Classifier Trained on Clean DS1 and Applied to Watermarked DS2 Without Retraining. (b) Bundled Ablation of the Physiological Objectives.}
\label{tab:fidelity_comparison}
\small
\setlength{\tabcolsep}{5pt}
\renewcommand{\arraystretch}{1.1}
\begin{tabular}{@{}l c c c c c c c c c@{}}
\multicolumn{10}{@{}l}{\textbf{(a) Comparison of watermarking methods}} \\
\toprule
\multirow{2}{*}{\textbf{Method}} & \multirow{2}{*}{\textbf{Train}} & \multirow{2}{*}{\textbf{Test}} & \multirow{2}{*}{\textbf{Acc.}} & \multirow{2}{*}{\textbf{Macro-F1}} & \multicolumn{5}{c}{\textbf{Sensitivity per AAMI class}} \\
\cmidrule(l){6-10}
 & & & & & \textbf{N} & \textbf{S} & \textbf{V} & \textbf{F} & \textbf{Q} \\
\midrule
No watermark (reference) & Clean & Clean & 90.24 & 35.25 & 95.10 & 8.89 & 71.97 & 0.29 & 0.00 \\
\midrule
ECGWmark (Ours) & Clean & Watermarked & \textbf{91.82} & \textbf{34.71} & \textbf{95.65} & 1.39 & \textbf{76.50} & 0.00 & 0.00 \\
LSB~\cite{halder2014patient} & Clean & Watermarked & \underline{89.42} & \underline{32.18} & \underline{93.80} & \textbf{4.12} & \underline{67.35} & 0.00 & 0.00 \\
DCT-SVD~\cite{jero2014dwt,anand2020improved} & Clean & Watermarked & 86.15 & 29.54 & 91.04 & 2.15 & 58.42 & 0.00 & 0.00 \\
SchurMark~\cite{tiwari2026schurmark} & Clean & Watermarked & 87.80 & 30.27 & 92.21 & \underline{2.68} & 61.26 & 0.00 & 0.00 \\
\bottomrule
\end{tabular}

\vspace{6pt}
\begin{tabular}{@{}l c c c c c c c c c@{}}
\multicolumn{10}{@{}l}{\textbf{(b) Ablation (ECGWmark, Clean$\rightarrow$Watermarked)}} \\
\toprule
\textbf{Configuration} & \textbf{Bit Acc.} & \textbf{SNR (dB)} & \textbf{Acc.} & \textbf{Macro-F1} & \textbf{N} & \textbf{S} & \textbf{V} & \textbf{F} & \textbf{Q} \\
\midrule
MSE reconstruction loss only & 55.84 & 10.21 & 87.36 & 26.66 & 92.98 & \textbf{5.28} & 71.87 & \textbf{0.29} & 0.00 \\
Full objective (Ours) & \textbf{95.27} & \textbf{12.00} & \textbf{91.82} & \textbf{34.71} & \textbf{95.65} & 1.39 & \textbf{76.50} & 0.00 & 0.00 \\
\bottomrule
\end{tabular}
\par\vspace{3pt}\parbox{\textwidth}{\footnotesize Bold and underlined values in (a) mark the best and second-best results among watermarked configurations, respectively; bold in (b) marks the better configuration. AAMI classes: N = normal and bundle branch block; S = supraventricular ectopic; V = ventricular ectopic; F = fusion; Q = unknown. The ablation modifies multiple objectives jointly and does not isolate $\mathcal{L}_{diag}$.}
\end{table*}

\subsection{Diagnostic Fidelity}
\label{sec:res_diag}
Table~\ref{tab:fidelity_comparison}(a) applies a fixed ResNet arrhythmia classifier, trained on clean DS1, to watermarked DS2 signals without retraining. Even on clean signals, this classifier rarely detects the rare classes (8.89\% sensitivity for Class S and almost 0\% for F and Q). Rare classes are known to be hard under the inter-patient protocol~\cite{dechazal2004automatic,luz2016ecg}, and our classifier was trained without class rebalancing, so macro-F1 and per-class sensitivity are more informative than overall accuracy. With ECGWmark, overall accuracy rises from 90.24\% to 91.82\% and Class-V sensitivity from 71.97\% to 76.50\%, but macro-F1 drops slightly from 35.25\% to 34.71\% and Class-S sensitivity drops from 8.89\% to 1.39\%. The effect on diagnosis is therefore mixed, and the loss of Class-S sensitivity is the main cost. Among the watermarking methods, ECGWmark has the highest accuracy, macro-F1, and Class-N and Class-V sensitivity, while LSB keeps the highest Class-S sensitivity (4.12\%). Because the baselines use different payloads and SNRs, this comparison is only descriptive.

\noindent\textbf{Independent classifier.} The classifier in Table~\ref{tab:fidelity_comparison}(a) is the same one that guides $\mathcal{L}_{diag}$, so testing only on it is partly circular. We therefore trained a separate 1-D ResNet with three random seeds and tested all four train$\rightarrow$test combinations, Clean$\rightarrow$Clean (C$\rightarrow$C), C$\rightarrow$W, W$\rightarrow$C, and W$\rightarrow$W, on DS2 with a patient-level bootstrap. The macro-F1 of C$\rightarrow$W was $+2.77\pm3.09$~pp higher than that of C$\rightarrow$C on average, but the sign changed across seeds and most per-seed confidence intervals included zero. The independent classifier thus shows neither a consistent loss nor a consistent gain, and each deployed classifier should be checked separately.

\subsection{Ablation of the Physiological Objectives}
\label{sec:res_ablation}
Table~\ref{tab:fidelity_comparison}(b) compares the full training objective against an MSE-only baseline ($\mathcal{L}_{rec}$). The full loss yields higher bit accuracy (95.27\% vs.\ 55.84\%), SNR (12.00 vs.\ 10.21~dB), and macro-F1 (34.71\% vs.\ 26.66\%), though lower Class-S sensitivity (1.39\% vs.\ 5.28\%). The MSE-only model stays near random chance in bit accuracy, showing the network fails to learn a usable watermark without spectral and diagnostic guidance. Because this ablation drops $\mathcal{L}_{diag}$ and $\mathcal{L}_{aux}$ together, it reflects their joint benefit rather than the isolated effect of $\mathcal{L}_{diag}$.

\subsection{Computational Cost and Visual Example}
\label{sec:res_cost}
The generator $G$ and detector $D$ have 313,377 and 247,970 parameters, totaling 561,347 (2.14~MiB in FP32) with 12.19~MiB peak GPU memory. For a 1,024-sample (2.84-s) segment, median runtime is 12.75~ms for embedding and 2.25~ms for extraction on an x86\_64 CPU (5.15~ms and 2.17~ms on an RTX~4090 GPU). The 15.00-ms total CPU time is just 0.53\% of the segment duration far faster than real time with negligible ECC overhead. While the model size matches wearable ECG networks~\cite{chen2023srecg,mao2026ecgpatch,rahman2024ecgnoise,liu2025wearableecg}, timings were measured on desktop hardware rather than microcontrollers.

\begin{figure}[t]
    \centering
    \includegraphics[width=0.8\linewidth]{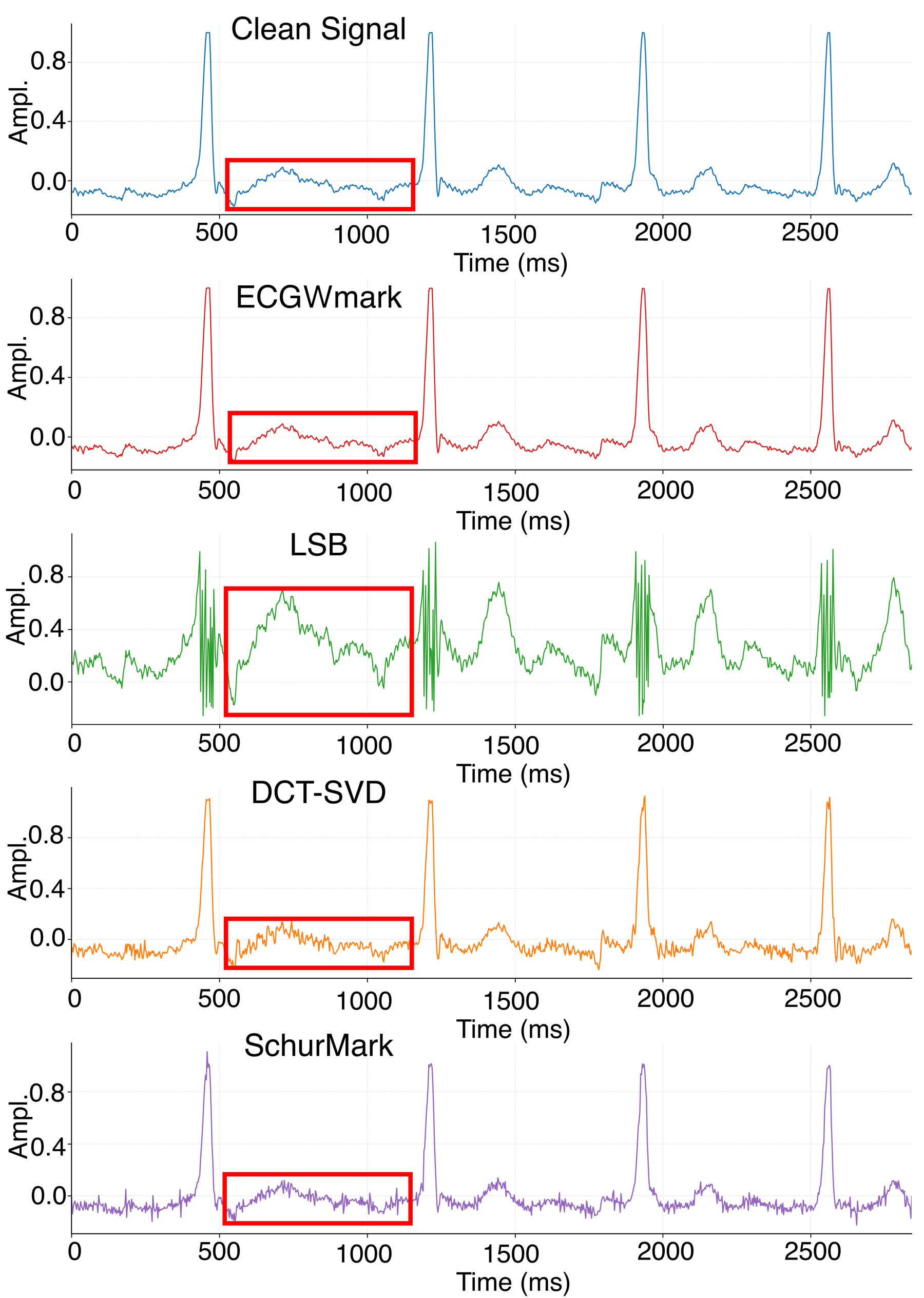}
    \caption{Clean and watermarked ECG segments across methods. Red boxes mark the low-amplitude interval between QRS complexes (T wave, baseline, and P wave) where artifacts are clearest, complementing Table~\ref{tab:fidelity_comparison}.}
    \label{fig:waveforms}
\end{figure}

Fig.~\ref{fig:waveforms} shows an example segment. LSB causes severe high-frequency oscillations near QRS complexes and distorts the baseline and T-wave peak, while DCT-SVD and SchurMark add visible baseline ripples. Conversely, ECGWmark closely follows the clean waveform, though single-segment visual similarity does not replace the quantitative diagnostic evaluation in Section~\ref{sec:res_diag}.

\section{Discussion}
\label{sec:discussion}
\subsection{Design Implications}
\label{sec:discussion}
\noindent\textbf{Robustness to time shifts versus robustness to noise.} The results show a consistent trade-off. Transform-domain methods spread each bit over a block of coefficients, which makes them robust to broadband noise and quantization, but the receiver must rebuild the same block grid. ECGWmark replaces this grid with local convolutional evidence averaged over time. It therefore tolerates shifts, filtering, and resampling, but keeps a small error on clean signals and is less accurate than DCT-SVD under moderate noise of 15--30~dB (Tables~\ref{tab:attacks_comparison} and~\ref{tab:matched}). In remote monitoring, segment offsets and dropped packets can occur and the original signal is not available at the receiver, so blind extraction that tolerates time shifts is the more useful property.

\noindent\textbf{Payload size, reliability, and choice of code.} Bit accuracy alone can be misleading. The same network that delivers a fully correct uncoded 32-bit message in only 14.10\% of clean segments delivers a fully correct 16-bit message in 99.86\% of them when half of the channels are used for redundancy. For patient identifiers or record pointers, a short message that is always correct is more useful than a longer one with frequent errors. We therefore recommend that learned watermarks carrying structured data report exact-message recovery together with the code rate. We chose the extended Hamming(8,4) code because exact ML decoding over 16 codewords is trivial; longer codes such as BCH codes~\cite{tancik2020stegastamp,lin2004errorcontrol} could carry more bits at the same reliability. Packet loss behaves more like missing bits (erasures) than like random bit errors, so erasure-aware decoding or an extra erasure code across consecutive segments could help. Adding a threshold on the decoding score in \eqref{eq:softml} would also allow unreliable segments to be rejected.

\noindent\textbf{Effect on diagnosis.} The results do not support the claim that the watermark leaves the diagnosis unchanged. For the guiding classifier, the watermark increased accuracy and Class-V sensitivity, slightly reduced macro-F1, and reduced Class-S sensitivity from 8.89\% to 1.39\%. Supraventricular ectopic beats are mainly recognized from early timing and small P-wave changes~\cite{luz2016ecg}, and the reference classifier already detects them poorly, so a small absolute drop is a large relative loss. Our experiments do not identify the cause of this drop. Possible remedies include a class-weighted $\mathcal{L}_{diag}$, a limit on residual energy near the P wave, and several guiding classifiers instead of one. Because the independent classifier gave results that varied across seeds and differed from those of the guiding classifier, the effect on diagnosis should be checked for every deployed model, and review by clinicians would be needed before clinical use.

\subsection{Limitations and Future Directions}
\label{sec:limitations}
\noindent\textbf{Cross-patient watermark training.} While our diagnostic classifier strictly follows the inter-patient DS1/DS2 split, the watermarking network was trained on randomly split segments. Because the encoder learns local, frame-level waveform residuals rather than patient-specific cardiac morphology, segment-level splitting is effective for learning general embedding rules; however, evaluating strictly patient-separated watermark training remains an important next step.

\noindent\textbf{Dataset scope and extreme channel loss.} Our evaluation focuses on fixed-length, single-lead segments from the MIT-BIH database. Extending ECGWmark to multi-lead systems, diverse sampling rates, and wearable recordings with heavy motion artifacts is a natural direction for future work. Furthermore, while our frame-level ECC handles moderate data drops well, severe 50\% packet loss reduces full-message recovery to 45.54\%. Overcoming such extreme link failures will require pairing ECGWmark with inter-segment erasure coding or retransmission protocols.

\noindent\textbf{Cryptographic security.} This study focuses on transmission robustness against unintentional channel distortions rather than adversarial security. For security-critical IoMT deployments, ECGWmark can be directly paired with lightweight encryption or cryptographic signatures prior to ECC encoding to prevent unauthorized removal or forgery.

\noindent\textbf{Granular ablations and edge deployment.} Our ablation study in Table~\ref{tab:fidelity_comparison}(b) evaluates the combined effect of our training objectives; isolating the individual contribution of $\mathcal{L}_{diag}$ will provide finer design insights in future analyses. In addition, while runtime was benchmarked on desktop hardware, the compact, fully convolutional design of ECGWmark makes it well suited for quantization and deployment on low-power wearable microcontrollers.

\section{Conclusion}
\label{sec:conclusion}
This article presented ECGWmark, a compact, quantization-free deep watermarking framework for single-lead ECG. Combining frame-level convolutions, a diagnostic loss, and soft-decision Hamming decoding, it jointly addresses time shifts, diagnostic safety, and full-message recovery. MIT-BIH experiments yield two key findings. First, frame-level extraction resists time shifts, filtering, resampling, and sample deletion—distortions that break block-transform methods. Second, while bit accuracy alone is insufficient, soft ECC decoding recovers full 16-bit messages in 99.86\% of clean segments and over 90\% under shifts, 10\% deletion, or 10\% packet loss. Thus, physiological watermarking should prioritize exact-message recovery and patient-independent diagnostic safety over waveform fidelity alone. Future work will extend ECGWmark to multi-lead signals and cryptographic provenance for IoMT monitoring. Code will be released upon publication.

\bibliographystyle{IEEEtran}
\bibliography{refs}

\end{document}